\documentstyle{l-aa-ps}     
\input{psfig}
 
\newcommand{\nxpup}{NX Pup}
\def\Ha{$\mbox{H}_\alpha$\/}

\begin{document}
 
   \thesaurus{06         % A&A Section 6: ??
              (08.09.2 NX Pup; 08.16.5;
               09.01.1;  % Image Processing
               03.20.3)  % Stars: individual
             }
   \title{Simultaneous optical speckle masking and NIR adaptive optics imaging 
of the 126\,mas Herbig Ae/Be binary star NX Puppis\thanks{Based
on observations obtained at the European Southern Observatory, La Silla}} 
%   \subtitle{}
   \author{Markus Sch\"oller \inst{1}, Wolfgang Brandner \inst{2}, Thomas Lehmann \inst{3}, Gerd Weigelt \inst{1}, \and Hans Zinnecker \inst{4}}
 
   \offprints{M. Sch\"oller}
 
   \institute{Max-Planck-Institut f\"ur Radioastronomie,
              Auf dem H\"ugel 69, D--53121 Bonn, Germany\\ ms@specklec.mpifr-bonn.mpg.de, weigelt@mpifr-bonn.mpg.de
\and
Astronomisches Institut der Universit\"at W\"urzburg, Am Hubland,
D--97074 W\"urzburg, Germany\\ brandner@astro.uni-wuerzburg.de
\and
Astrophysikalisches Institut und Universit{\"a}ts-Sternwarte Jena,
Schillerg{\"a}{\ss}chen 2, D--07740 Jena, Germany\\ lehmi@sol.astro.uni-jena.de
\and 
Astrophysikalisches Institut Postdam, An der Sternwarte 16, D--14882 Potsdam,
Germany\\
hzinnecker@aip.de}
 
\date{Received date; accepted date}
 
\maketitle

\markboth{Sch\"oller et al.:}{Simultaneous speckle masking and adaptive optics 
imaging of NX Puppis}
 
\begin{abstract}
We present simultaneous optical and near-infrared high angular resolution 
observations of the close Herbig Ae/Be binary star NX Pup which
is associated with cometary globule~1. 
The reconstructed images have a diffraction--limited resolution of 62\,mas in V,
75\,mas in R (speckle masking reconstruction), and 115\,mas in H,
156\,mas in K (adaptive
optics + post--processing).
Compared to previous results we were able to derive 
better estimates on spectral type and luminosity and hence put better
constraints on the evolutionary status (mass \& age) of NX Pup A and B:
with NX Pup A of spectral type F0-F2 we estimate the spectral type of
NX Pup B in the range F7-G4, masses of 2 M$_\odot$ and 1.6--1.9 M$_\odot$,
respectively, and an age of 3--5~Myr for both stars.

We discuss the implication of the new age determination on the physical
relation between NX Pup and the cometary globule. The dynamical lifetime
of $\approx$\,10$^6$ yr for cometary globule 1 suggests that cometary globule 1
and the nearby cometary globule 2 represent transient phenomena and are 
left overs of a larger molecular cloud which in turn was the parental cloud
of NX~Pup A and B and finally got dispersed by photoevaporation.

The IR excess of NX Pup A can be modeled by a viscous accretion disk, which
is cut off at $\approx$\,20 AU from the star. NX Pup B has a smaller
IR excess which indicates that there is less circumstellar material present
than around the primary.
\end{abstract} 
\section{Introduction}
\begin{figure}[ht]
\centerline{\vbox{\psfig{figure=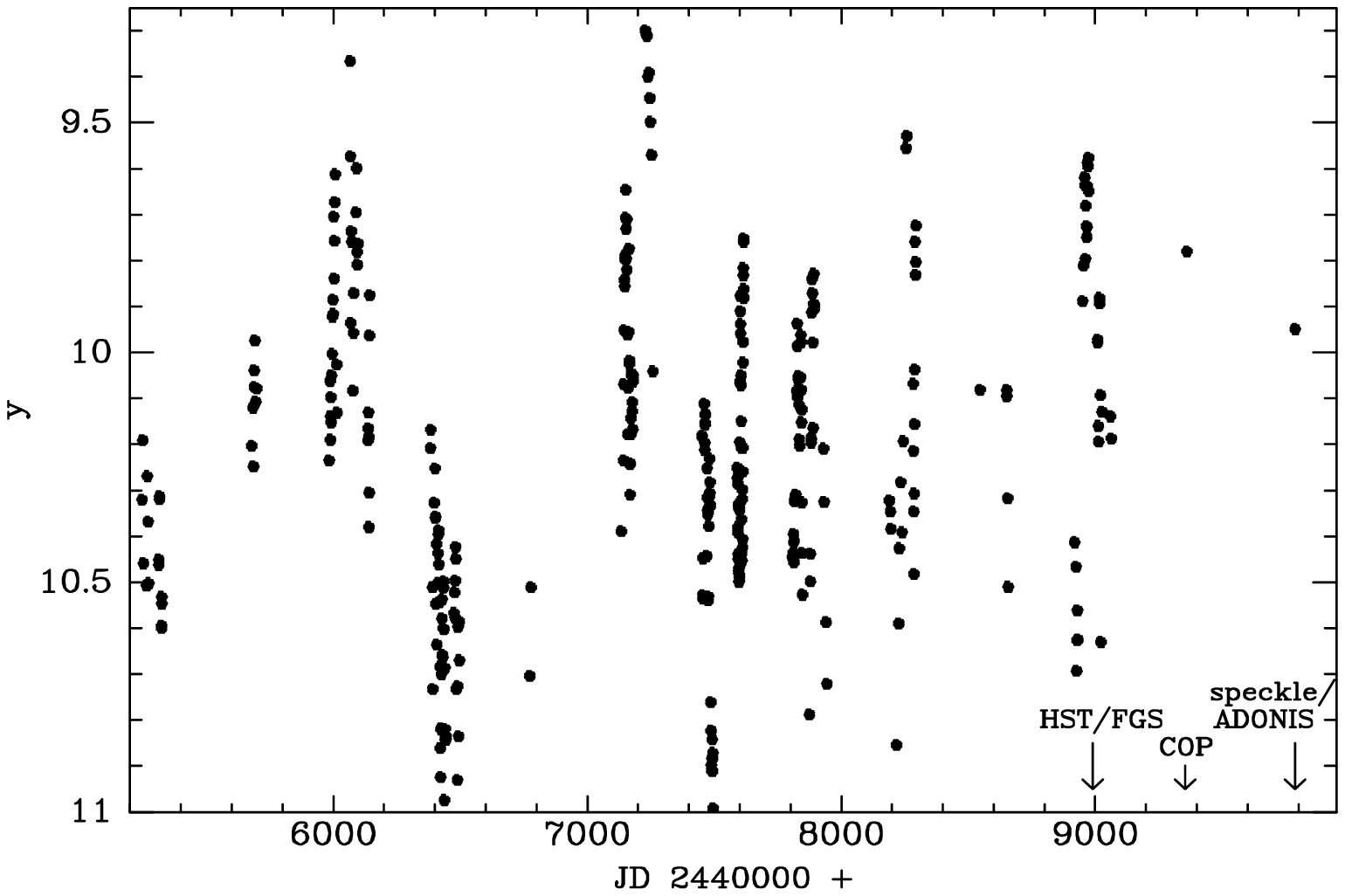,width=8.5cm,height=5.5cm}
                  \psfig{figure=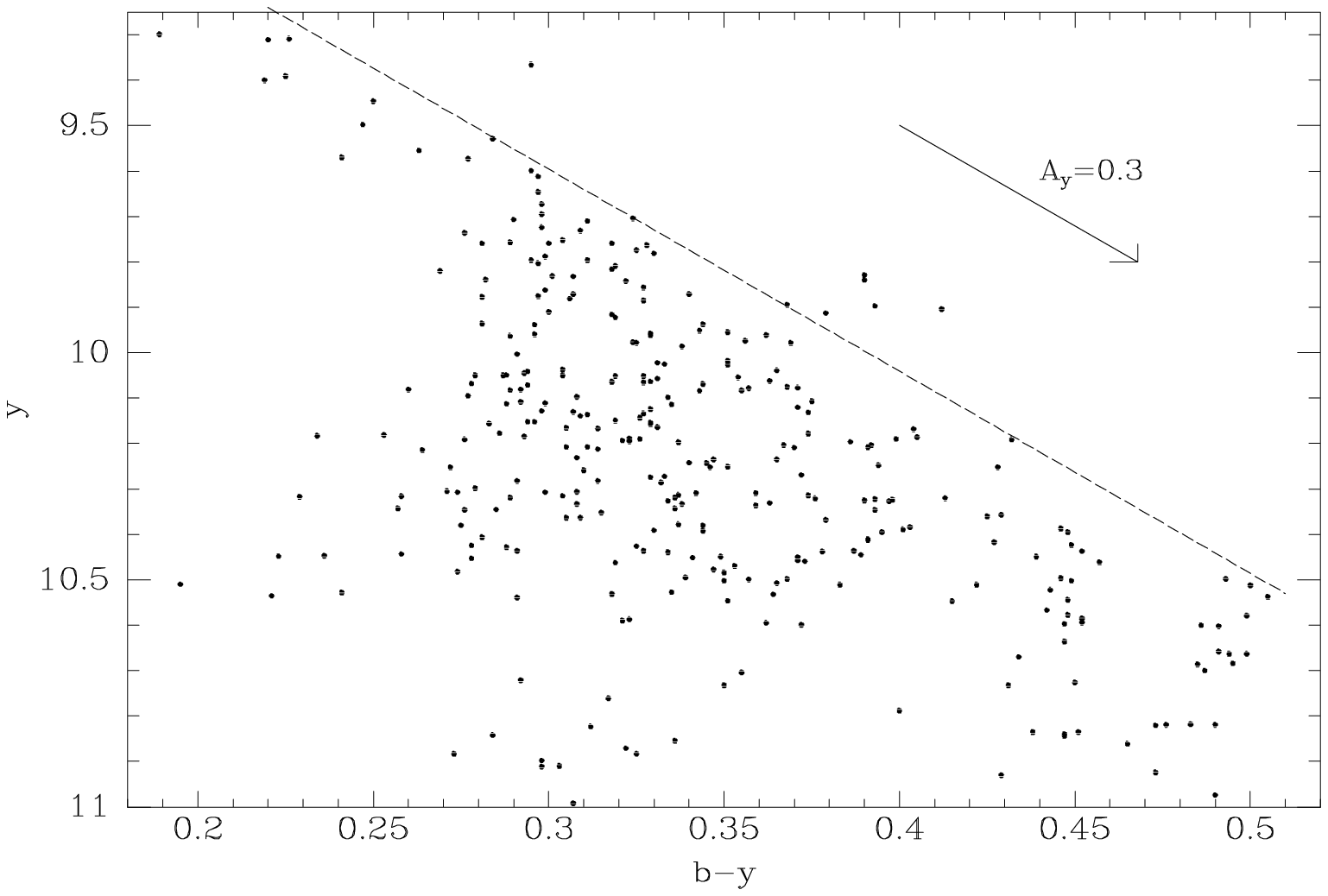,width=8.5cm,height=5.5cm}}}
\caption{\label{v_jd} y (in the Str\"omgren photometric system) light curve 
(top) and corresponding y vs.\ b--y
CMD (bottom) of NX Pup from 1983 to 1995 (compiled from LTPV data).
Note the rapid variations with an amplitude of $\approx$\,1\fm7.
Marked by an arrow are the dates of the HST/FGS observations, the first
NIR adaptive optics observations (COP), and the current simultaneous
data sets presented in this paper. At the time of all three high angular 
resolution observations NX Pup was relatively bright.
The CMD shows no clear colour--magnitude relation. However, the envelope
(dashed line) is parallel to the reddening arrow (assuming a standard
interstellar extinction law), which indicates that 
variable extinction is an important constituent of the overall brightness
variations.}
\end{figure}

NX Puppis is a Herbig Ae/Be star (Irvine 1975) associated with
cometary globule 1 (CG1, e.g.\ Harju et al.\ 1990) which itself
is located in the Gum nebula at a distance of $\approx$\,450 pc (Smith 1968,
Brandt et al.\ 1971, Herbst 1977). 
NX Pup shows a strong 
UV excess (de Boer 1977) and IR excess (e.g.\ Brand et al.\ 1983). 
Hoffmeister (1949) was the first to notice it as a highly variable star.
The photometric variability has been studied extensively by Bibo \& Th\'e
(1991).
They found NX Pup to be unique in the sense that the dependence of
colour index on magnitude could neither be described as monotonic
nor non-monotonic.
In Fig.\ \ref{v_jd} we have plotted the y (in the Str\"omgren photometric
system) light curve of NX Pup from 
1983 till 1995 and the corresponding colour--magnitude diagram,
compiled from LTPV data (Long-Term Photometry of Variables Project,
Manfroid et 
al.\ 1991 \& 1995; Sterken et al.\ 1993 \& 1995) and our own data.

The binary nature of NX Pup was first revealed by observations with the Fine
Guidance Sensors (FGS) aboard the Hubble Space Telescope (HST)
(Bernacca et al.\ 1993). Follow-up observations of this close
Herbig Ae/Be binary (126mas, $\approx$\,60 AU) were carried out by Brandner
et al.\ (1995) using adaptive optics in the near-infrared. They found that
the two components NX Pup A \& B are very likely pre-main sequence stars 
both exhibiting an IR excess. Furthermore, they assigned for
component A a spectral type A7--F2, an age of $5 \times 10^6$ yr,
and a mass of $\approx$\,2\,M$_\odot$, whereas the physical parameters
for component B were more uncertain because of the lack of high
angular resolution data in the optical.
The spectral type of NX Pup B lies in the range of F5 to G8, the age is
between 0.3 and $5 \times 10^6$ yr, and the mass is between
$\approx$\,1.5\,M$_\odot$ and 2.5\,M$_\odot$.

The age and mass determinations for both components were also
uncertain because of the variability of NX Pup and the time gap between
the different sets of observations: the adaptive optics JHK observations
were carried out exactly one year after the 550\,nm (``V'')
HST/FGS observations (cf.\ Fig.\ \ref{v_jd}). Furthermore, the HST observation
provided only
the brightness difference between the two components but no absolute
calibration. The brightness of NX Pup at the time
of the HST observation could only be estimated by interpolation of
photometric measurements before and after the HST observation.

To remedy this situation, we initiated simultaneous optical and NIR
high spatial resolution observations of NX Pup with the aim of obtaining
estimates for the binary components in order to put better constraints
on their evolutionary status and to probe the circumstellar matter.
Furthermore we intended to find clues about the relation of NX Pup to CG1.

All optical and infrared data were obtained within 3\,hr between 00:15 and
03:15 UT on Mar 11 1995 (cf.\ Table \ref{obs}) which ensures that the 
variability of NX Pup should affect our conclusions to only a small extent.

\section{Observations and data reduction}
\begin{table*}
\caption{\label{obs}Journal of observations}  
\begin{center}
\begin{tabular}{lllc}
telescope/instrument& date (UT) & filter ($\lambda_c$, FWHM) & exposure time\\ \hline
D1.54/CCD-camera     &11 March 1995 (01:30) & B,V,R& 5s, 2s, 1.5s    \\
ESO/MPG 2.2m/speckle camera &11 March 1995 (01:15) & ``V'' (545nm, 30nm) &629$\times$ 50ms\\   
ESO/MPG 2.2m/speckle camera &11 March 1995 (00:15, 03:00) & ``R'' (656nm, 60nm) &1927$\times$ 70ms\\  
ESO/MPG 2.2m/speckle camera &11 March 1995 (01:00, 02:45) & ``R'' (656nm, 30nm) &2285$\times$ 70ms\\  
ESO/MPG 2.2m/speckle camera &11 March 1995 (03:15) & \Ha\ (656.3nm, 4nm) &1903$\times$ 70ms\\   
ESO3.6m/ADONIS+SHARP&11 March 1995 (02:30) & H, K & 400$\times$ 0.5s (each)\\   
\hline
\end{tabular}
\end{center}
\end{table*}

\subsection{Speckle masking results}
The \nxpup{} speckle data were obtained with the ESO/MPG 2.2m telescope 
at La Silla on March 10/11, 1995.
Table \ref{obs} gives an overview of the four speckle data sets which were
recorded.
The FWHM diameter of the motion-compensated
long-exposure image was about 0.5$''$.

The speckle raw data were recorded with the speckle
camera described by Baier \& Weigelt (1983).
The detector used for the observations was an
image intensifier (gain 500~000, quantum efficiency
about 10\,\% at 545~nm and about 8\,\% at 656~nm) coupled optically to a
fast CCD camera ($512^2$ pixels/frame, frame rate
4 frames/sec, digital correlated double sampling).
A system of Digital Signal Processors was used for real-time speckle interferometry
and real-time speckle masking, and for fast simultaneous data storage on four Exabyte streamers.

\begin{figure}[ht]
%\centerline{\psfig{figure=fig2.ps,width=8.5cm,height=8.5cm}}
\caption{\label{ImaI}
Diffraction--limited V, R, and \Ha\ speckle masking reconstructions
and H--band adaptive optics image of NX Pup.
The faint pattern north of NX Pup in the \Ha-- and H--band image very likely
is an artefact.}
\end{figure}

\begin{figure}[ht]
%\centerline{\vbox{\psfig{figure=fig3a.ps,width=8.5cm,height=5.5cm}
%                  \psfig{figure=fig3b.ps,width=8.5cm,height=5.5cm}}}
\caption{\label{ImaIII}
Surface plot of the R (FWHM 30 nm, top) and \Ha\ (FWHM 4\,nm, bottom) speckle
masking reconstruction of NX Pup.
The intensity ratio is $\approx$\,0.620 in the R image and $\approx$\,0.375
in the \Ha\ image.
The images show that even in the 4\,nm \Ha\ reconstruction the height
of the strongest noise peaks is much smaller than the difference
of the R and \Ha\ intensity ratio.
}
\end{figure}

A diffraction-limited image (cf.\ Fig.\ \ref{ImaI}) of \nxpup{} was 
reconstructed from the speckle interferograms by the speckle masking
method
(Weigelt 1977; Lohmann et al.\ 1983;
Hofmann \& Weigelt 1988).
The following processing steps were applied to each of the four speckle
data sets:

\begin{enumerate}
\item Subtraction of the average CCD noise bias and division by the flatfield
for each speckle interferogram
\item Calculation of the average power spectrum of all speckle data
\item Subtraction of detector noise bias terms from the average power spectrum
\item Compensation of the photon bias terms in the average power spectrum
\item Calculation of the average bispectrum of all speckle interferograms
\item Subtraction of detector noise bias terms from the average bispectrum
\item Compensation of the photon bias terms in the average bispectrum
\item Compensation of the speckle interferometry transfer function in the 
bias-compensated average power spectrum to obtain the Fourier modulus (Labeyrie 1970)
\item Retrieval of Fourier phase from the bias-compensated average bispectrum
\item Reconstruction of the diffraction-limited image from the object modulus
and phase
\end{enumerate}

The photon bias terms in the average power spectrum and bispectrum
(due to the spatially extended photoevents of our image intensifier)
were compensated by the method described by Pehlemann et al.\ (1992).
The object Fourier phase was reconstructed from the bias-compensated
average bispectrum using the conventional phase recursion method
(Lohmann et al.\ 1983).

The bispectrum of each frame consisted of $\approx$\,35 Million elements
(maximum length of bispectrum vectors: maximum u-vector 50 pixels, maximum
v-vector 98 pixels, diffraction
cut-off frequency at pixel 98) for the R and \Ha\ filter data sets.
For the V filter data set the bispectrum consisted of $\approx$\,59 Million elements
(maximum length of bispectrum vectors: maximum u-vector 55 pixels,
maximum v-vector 113 pixels, diffraction cut-off frequency at pixel 113).
Each bispectrum element was weighted with its signal--to--noise ratio (SNR) 
for correct weighting in the phase recursion algorithm.

Figures \ref{ImaI} and \ref{ImaIII} show the results of our speckle observations
of \nxpup{}.
No postprocessing by an image restoration
method was applied to the speckle masking reconstructions.
The reconstructed images have diffraction-limited resolution of 62\,mas in V
and 75\,mas in R.

\begin{table}
\caption{\label{spa}\nxpup{} parameters derived from the speckle masking observations at epoch 1995.19}
\begin{center}
\begin{tabular}{l|lll}
filter ($\lambda_c$/FWHM) & Separation & PA & Intensity ratio \\
\hline
narrow V (545/30)& 128 $\pm$ 3 mas & 62\fdg1 $\pm$ 1\fdg7 & 0.537 $\pm$ 0.030 \\
broad R (656/60) & 124 $\pm$ 3 mas & 63\fdg8 $\pm$ 1\fdg7 & 0.620 $\pm$ 0.025 \\
narrow R (656/30)& 125 $\pm$ 3 mas & 62\fdg6 $\pm$ 1\fdg7 & 0.625 $\pm$ 0.020 \\
\Ha\ (656/4)     & 122 $\pm$ 3 mas & 61\fdg1 $\pm$ 1\fdg7 & 0.375 $\pm$ 0.040 \\
\hline
\end{tabular}
\end{center}
\end{table}

A surface plot of the R and \Ha\ image is shown in Fig.\ \ref{ImaIII}.
NX Pup A is significantly brighter than NX Pup B in \Ha.
The intensity ratio, separation, and position angle (PA) of the two components
of \nxpup{} were determined using the IRAF package APPHOT
(cf.\ Table \ref{spa}).

\subsection{Photometric calibration}
CCD photometry of the unresolved pair NX Pup A/B was carried out at the Danish
1.5m telescope at La Silla. Observations of standard stars taken from the list
by Landolt (1992) allowed for the absolute photometric calibration.

\begin{table}
\caption{\label{VRJHK} Optical and NIR photometry (11 March 1995) and IRAS
measurements for component A and B. If only one value is given, it corresponds
to the combined (unresolved) brightness/flux of both components.}
\begin{center}
\begin{minipage}{70mm}
\begin{tabular}{cccc}
filter& NX Pup A & NX Pup B   \\
\hline
B       & \multicolumn{2}{c}{10\fm35$\pm$0\fm02 }\\
V       & 10\fm43$\pm$0\fm05 & 11\fm11$\pm$0\fm05  \\
R       & 10\fm13$\pm$0\fm05 & 10\fm64$\pm$0\fm05  \\
H$\alpha$\footnote{See discussion of
the problem concerning the uncertainties in calibrating the \Ha\ filter in the text.}
        & 8\fm98$\pm$0\fm10 & 10\fm04$\pm$0\fm10 \\
J\footnote{1.1.1994 (Brandner et al.\ 1995)}  & 8\fm58$\pm$0\fm05 & 9\fm56$\pm0$
\fm05 \\
H       & 7\fm47$\pm$0\fm05 & 8\fm26$\pm$0\fm05  \\
K       & 6\fm34$\pm$0\fm05 & 7\fm66$\pm$0\fm05 \\ \hline
12$\mu$m& \multicolumn{2}{c}{7$\pm$0.5 Jy}\\
25$\mu$m& \multicolumn{2}{c}{10$\pm$1 Jy}\\
60$\mu$m& \multicolumn{2}{c}{24$\pm$2 Jy}\\
100$\mu$m& \multicolumn{2}{c}{70$\pm$5 Jy}\\ \hline
\end{tabular}
\end{minipage}
\end{center}
\end{table}

The 545 nm and 656 nm continuum filters used for the speckle observations
are somewhat distinct from the Bessell V and Cousin R filters used in our
CCD observations.
The speckle ``V'' filter has the same central wavelength but is about 4 times 
narrower than the standard filter. For the speckle ``R'' filter the central
wavelength is off by about 10 nm. Hence, we expect systematic errors due to 
this mismatch of passbands in V smaller than 0.01 mag and in R up to 0.05 mag 
(depending on the spectral type of the observed object). 
This effect, however, will be neglected in the following.
For the \Ha\ speckle images no direct photometric calibration was available.
However, the ratio of the passbands of the narrowband \Ha\ filter 656/4 and the 
broadband R filter 656/60 is 1:13.5. While for the \Ha\ observations
the stop number of the coupling optics behind the image intensifier was set to 2.0, it was
set to 2.8 for the R observations. Hence we get an efficiency ratio
of 1:7 between the \Ha\ and the broadband R observations, which is also
confirmed by analysing flatfield data.
The \Ha\ brightness can be defined by postulating that stars without
\Ha\ emission or absorption should have R--\Ha = 0. 
NX Pup appears to be about 0.4 mag brighter in \Ha\ compared
to the \Ha\ brightness observed by Brandner et al.\ (1995).
This can be explained by a brightening of the \Ha\ emission from NX Pup.
However, it is still possible that our indirect calibration of the
\Ha\ brightness is not perfect.
So far we can only say that both stars show \Ha\ in emission,
that the majority of the \Ha\ excess
originates in NX Pup A, and that this excess is possibly related to
the circumstellar matter present around NX~Pup A.

All photometric reductions were done using IRAF. The results are summarized in
Table \ref{VRJHK}.

\subsection{Adaptive optics imaging}

The adaptive optics data were obtained at the ESO 3.6\,m using ADONIS (ADaptive 
Optics Near Infrared System, e.g., Beuzit et al.\ 1994) in combination with
the SHARP (System for High Angular Resolution Pictures, e.g., Hofmann et al.\ 1995) camera, operated
by the Max Planck Institute for Extraterrestrial Physics. NX Pup was observed
in the H and K band with a total integration time of 200\,s each.
Immediately after NX Pup a nearby single star with similar brightness and
spectral characteristics was observed as a point source reference.

The data were sky subtracted and flat fielded (using dome flats) following
the standard reduction procedure.
Stars out of the ISO list of photometric standards (Bouchet,
private comm.) were observed several times throughout the night for
extinction determinations and absolute photometric calibrations.

The close binary (sep.\ = 0\farcs126) is already clearly resolved on the
best single exposures. 
In order to improve the spatial resolution we selected those 10\,\%
of the exposures with the highest Strehl ratio (see also the discussion
in Brandner et al.\ 1995) and coadded them.

As adaptive optics systems only partially compensate the wavefront
distortions, the uncompensated power forms a halo around each object.
However, at the time of the observations ADONIS did not allow for a direct
(simultaneous to the scientific observations) determination of the 
(instrumental) point spread function (PSF).
Therefore a single star had to be observed before and after
the scientific observations in order to obtain a PSF calibration. 
In order to remove the uncompensated halo, the same selection criteria that 
we used for NX Pup were also applied to the PSF star. The instrumental signature
was then (partially) removed by applying two iterations of a ``simple
deconvolution'' (Blecha \& Richard 1989). The achieved resolution is
115 mas in H and 154 mas in K.
Figure \ref{ImaI} (bottom, right) shows the H image
of NX Pup.
The great
drawback of this method is that -- as observing conditions are changing --
the PSF itself is varying and hence no perfect compensation and removal
of the instrumental signatures is possible (cf.\ Fig.\ \ref{ImaI}). 
The resulting H and K magnitudes can be found in Table \ref{VRJHK}.

\section{Separation, PA, and photometric variability}

By averaging the results of the three V and R filter speckle observations we
obtain a separation of 126 mas $\pm$ 3 mas and a PA of 62\fdg8 $\pm$ 1\fdg7.
These values are in good agreement with previous results obtained by Bernacca
et al.\ (1993, epoch 1993.0: sep.\ = 126 $\pm$ 7 mas, PA = 63\fdg4 $\pm$ 1\fdg0) and
Brandner et al.\ (1995, epoch 1994.0: sep.\ = 128 $\pm$ 8 mas, PA = 62\fdg4 $\pm$ 5\fdg7).

Due to the relatively large uncertainty in the determinations of the PA, we
still see no evidence for orbital motion. Assuming a system
mass of 4 M$_\odot$, a semimajor axis of 60 AU (126 mas at 450 pc), 
and a circular orbit
perpendicular to the line of sight we would expect a shift in PA by 
$\approx$\,1\fdg5/yr. The lack of such a shift would either
mean that our assumptions with respect to the orbital parameters are wrong
(i.e.\ inclination $\neq$ 0$^\circ$ or e $>$ 0) or that the distance to
NX Pup is considerably larger than 450 pc. Here HIPPARCOS observations --
once they are made available -- should help to clarify the situation.

Before trying to decompose the spectral energy distribution of NX Pup
we have to evaluate what might be the reason for its variability.
Different scenarios have been proposed to explain brightness variations 
in Herbig Ae/Be stars and T Tauri stars: (i) star spots, 
detected first by Bouvier \& Bertout (1989), which can explain 
quasi--periodic variations observed in weak--line T Tauri stars
(ii) solar--type 
(albeit on a much larger scale) flare
events, which can explain the sudden rise in brightness, observed
for weak--line T Tauri stars in the optical and in X--rays 
(e.g.\ Gahm et al.\ 1995, Guenther 1995, Preibisch et al.\ 1995), 
(iii) variations in the veiling continuum/accretion luminosity,
which cause irregular brightness variations in classical T Tauri stars
and outburst phenomena like those observed in FUORs and EXORs (e.g.\ Lehmann
et al.\ 1995), and (iv) variable obscuration by circumstellar material
(Grinin 1992) or protoplanets forming in the plane of a circumstellar
disk (Th\'e \& Molster 1994).

\begin{table}
\caption{\label{var} Amplitude of variability of NX Pup}
\begin{center}
\begin{tabular}{cccc}
filter& range& $\Delta$M & ref.\  \\
\hline
u       & 10\fm74--13\fm09 & 2\fm3 &LTPV   \\
v       &  9\fm79--11\fm97 & 2\fm2 &LTPV   \\
b       &  9\fm49--11\fm46 & 2\fm0 &LTPV   \\
y       &  9\fm30--10\fm99 & 1\fm7 &LTPV   \\
R       &  9\fm46--10\fm73 & 1\fm3 &BHLWC, R   \\
I       &  9\fm01--10\fm28 & 1\fm2 &BHLWC, R   \\
J       &  8\fm01--8\fm49  & 0\fm5&BHLWC, R    \\
H       &  6\fm94--7\fm24  & 0\fm3&BHLWC, R    \\
K       &  5\fm85--6\fm06  & 0\fm2&BHLWC, R    \\
L       &  4\fm40--4\fm50  & 0\fm1&BHLWC, R    \\
\end{tabular}
\end{center}
References: BHLWC - Brand et al.\ 1983, R - Reipurth 1983,
 LTPV - Sterken et al.\ 1995
\end{table}

Recently, Eaton \& Herbst (1995) studied UV spectra of 5 UXOR type Herbig
Ae/Be stars and gathered evidence that the dominant source of variability
is variable extinction. They argue that protoplanets are unlikely responsible 
for the observed obscuration, because this would require nearly edge--on 
viewing of the orbits for {\it all} systems which exhibit this kind of 
variability.

The general change in colour with brightness is intriguing: the envelope of
the CMD in Fig.\ref{v_jd} is parallel to the reddening vector, 
and hence brightness--colour variations along this line can be explained 
by variable circumstellar extinction. 
However, near minimum light NX Pup can be as blue as in maximum
light. We interpret this as clear evidence for a colour reversal
(``blueing effect'', see Wenzel 1969)
in some observing periods, especially when NX Pup
is faint. The large range of colours of NX Pup at low light level may indicate
different strengths of the blueing effect in different observing periods.

By analyzing the lightcurve of NX Pup in short time intervals
of good coverage we find the following distinct behaviours: (i) changes in
magnitude and colour parallel to the reddening vector when the star is bright, 
(ii) fast quasiperiodic oscillations (with an amplitude of about 0.6\,mag, 
typical period 14\,d) without correlated colour variation, observed when 
the star exhibits mean or low brightness, (iii) periods of small light 
variation 
with little colour change, and (iv) sudden decreases in brightness without any
colour change between two periods of small variability.

In general, the timescale of colour variations is much longer than the
timescale of intensity variations.
Many of the features visible in the brightness and colour of NX Pup have been
recorded for other Herbig Ae/Be stars (e.g.\ HR\,5999, Th\'e 1994) or T Tauri 
stars (e.g.\ RY Lup, Gahm et al.\ 1989,1993).
However, NX Pup is a binary star as well. If we assume that component NX
Pup B would not be variable at all, the amplitude of brightness
variations  for NX Pup A would become considerably larger than the amplitude 
measured for the unresolved binary system (see Table \ref{var}). 
At low luminosity,
also the spread in colours for NX Pup A would increase even more.

The fact that no veiling of photospheric lines has been reported in recent high
resolution spectroscopy (B\"ohm \& Catala 1995) supports the variable 
obscuration scenario rather than models of variable accretion. 
As, in general, circumstellar extinction dims and reddens the light of a 
star, we argue -- following the reasoning of Gahm et al.\ (1989) --
that the spread in colours at any brightness level (and especially 
the blueing of NX Pup when it becomes faint) requires an additional source of 
variability not affected by the obscuration and hence being of circumstellar 
nature. We propose that the light seen from NX Pup near
minimum brightness is mainly scattered light from a circumstellar disk or
envelope, while the direct light from the star is heavily obscured.
The large range of possible colours at low brightness level of NX Pup
could be the result of a variable amount and/or distribution of the scattering
particles. The photometric data reveal that the timescale of variability
due to scattered light is of the order of years, whereas occultations and 
associated quasiperiodic variability occur on timescales of days.

\section{SED and the evolutionary status of NX Pup A \& B and CG1}

\begin{figure*}[ht]
\centerline{\psfig{figure=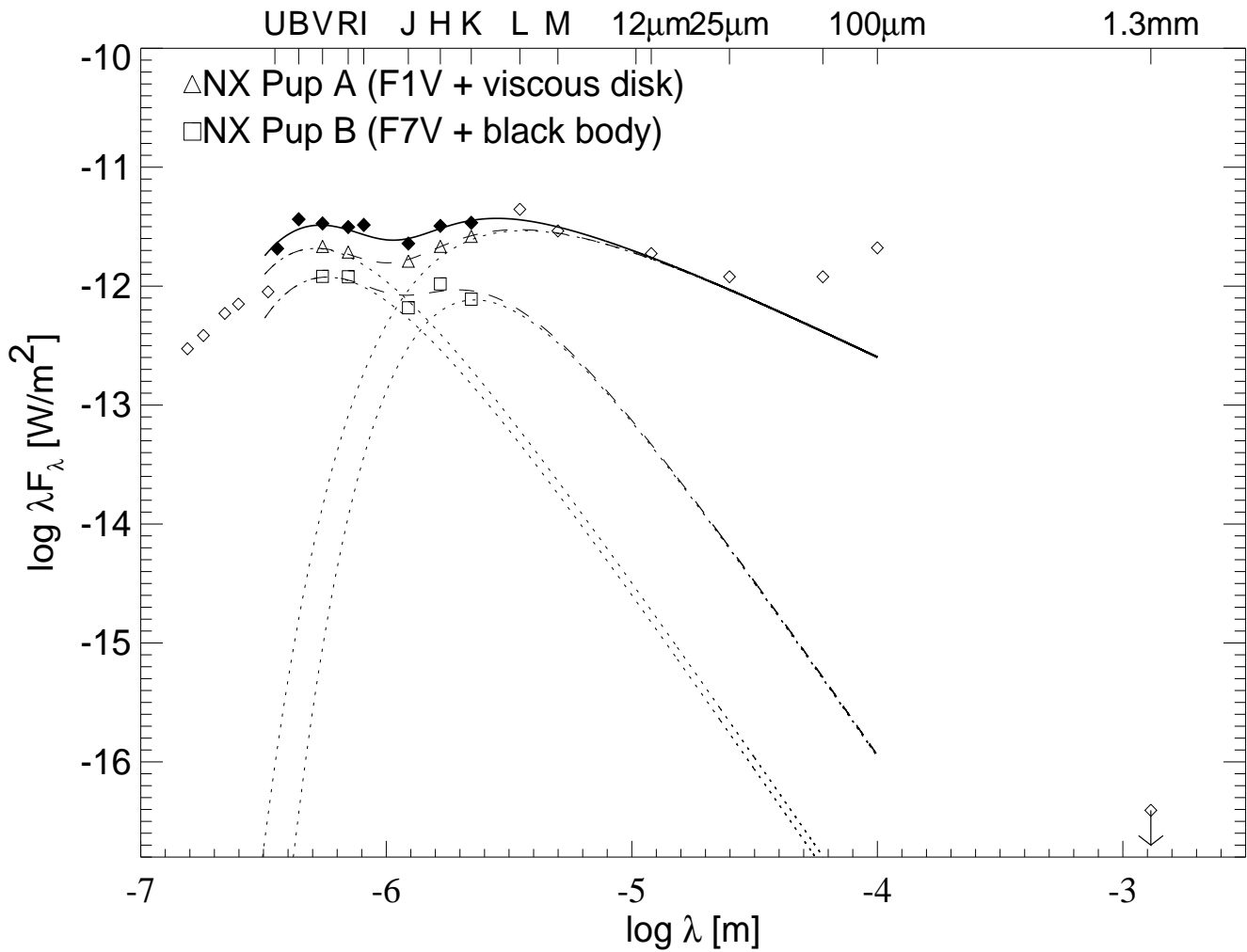,width=19cm,height=14cm}}
\caption{\label{sed} Dereddened (assuming A$_v$ = 0\fm48) spectral energy
distribution $\lambda$F$_\lambda$ of NX Pup A and B from the UV to the
mm-range. The filled diamonds and open triangles and squares
indicate our own measurements.
The lack of strong 1.3 mm emission suggests
that the circumstellar disk around NX Pup A is cut off at about 20 AU and
that there is no circumbinary disk present around NX Pup A/B.
The total spectral energy distribution can be decomposed into four parts (dotted lines):
the photospheric emission from NX Pup A (F1V) and NX Pup B
(F7V), a viscous accretion disk around NX Pup A, and circumstellar
matter around NX Pup B which is approximated by a blackbody.
The dashed lines mark both the SEDs of the individual stellar photospheres plus
the IR excess due to circumstellar matter for NX Pup A and B.
The overall SED is indicated by a solid line and gives a reasonable fit
to the observed flux distribution (diamonds).
The errors in flux are 5\,\% or less (see text for more details).}
\end{figure*}

Superimposed on the variable obscuration effects, 
varying accretion luminosity may explain some of the 
scatter in the CMD. Using IUE spectra (120 nm -- 330 nm) of NX Pup, 
Blondel \& Tjin A Djie
(1994) could show that the UV spectrum can be decomposed into an underlying
stellar photosphere plus accretion luminosity and a viscous accretion disk.
Their model consists of a star with a spectral type F2 III--V, a boundary 
layer with a temperature of $\approx\,10^4$ K (resembling the spectrum of an 
A0 supergiant) and a cool disk. 
The adopted visual extinction was 0\fm4 to 0\fm8.

Similar to the decomposition of the UV SED we can try to decompose the
optical and infrared SED of NX Pup A \& B.
Figure \ref{sed} shows the overall spectral energy distribution of
NX Pup from the UV to the mm--range.
The UV values are from de Boer (1977, epoch: 1975.83), the U to K from our own measurements
(this paper \& Brandner et al.\ 1995), the L and M values from
Hillenbrand et al.\ 1992, the 12 $\mu$m to 100 $\mu$m IRAS fluxes
(cf.\ Table \ref{VRJHK}) were measured on FRESCO/HIRES images obtained
from the Infrared Processing \& Analysis Center (IPAC)
\footnote{IPAC data can be retrieved on the WWW via
http://www.ipac.caltech.edu/}, and the
1.3~mm value, which represents an upper limit, is from Henning et al.\ (1994).

Most of the light detected in V and R is emitted by the stellar 
photospheres of NX Pup A and B. Both components also
exhibit a NIR excess. While the SED of NX Pup B peaks at H, the SED of NX Pup
A is still rising at K. Hence, the majority of the IR excess arises from
component A.  The lack of strong emission from cold dust at 
1.3 mm, which mainly originates from the cooler outer parts of the disk,
suggests that the inner disk around NX Pup A is cut off at
about 20 AU and that there is no massive circumbinary disk present.
Theoretical calculations of disk sizes (both circumstellar
and circumbinary) in binary systems by Papaloizou \& Pringle (1977)
and Artymowicz \& Lubow (1994) indicate that the circumprimary
disk may extend out to 0.4 times the semimajor axis of the binary
orbit, which for NX Pup is about 20 AU, whereas the circumsecondary disk should be
considerably smaller (about 10 AU in our case). The predictions,
however, are sensitive to parameters like mass ratio, eccentricity of binary
orbit, or viscosity parameter $\alpha$. 

Based on low to intermediate resolution spectroscopy,
Brand et al.\ (1983) and Reipurth (1983) classified NX Pup as F0--F2.
Using the spectral type -- effective temperature calibration
from de Jager \& Nieuwenhuijzen (1987) and the colours and bolometric
corrections as tabulated by Hartigan et al.\ (1994) we are able
to place NX Pup A in an H-R diagram.
If we adopt a MK type of F1 IV--V for NX Pup A, the observed V--R colour
(0\fm30) yields a visual extinction A$_V \approx$\,0\fm48. This value
is in good agreement with the value derived by Blondel \& Tjin A Djie
(1994) from best fits of IUE spectra of NX Pup (A$_v \approx 0\fm43$)
and allows us to determine L$_{bol}$.

\begin{figure}[ht]
\centerline{\vbox{\psfig{figure=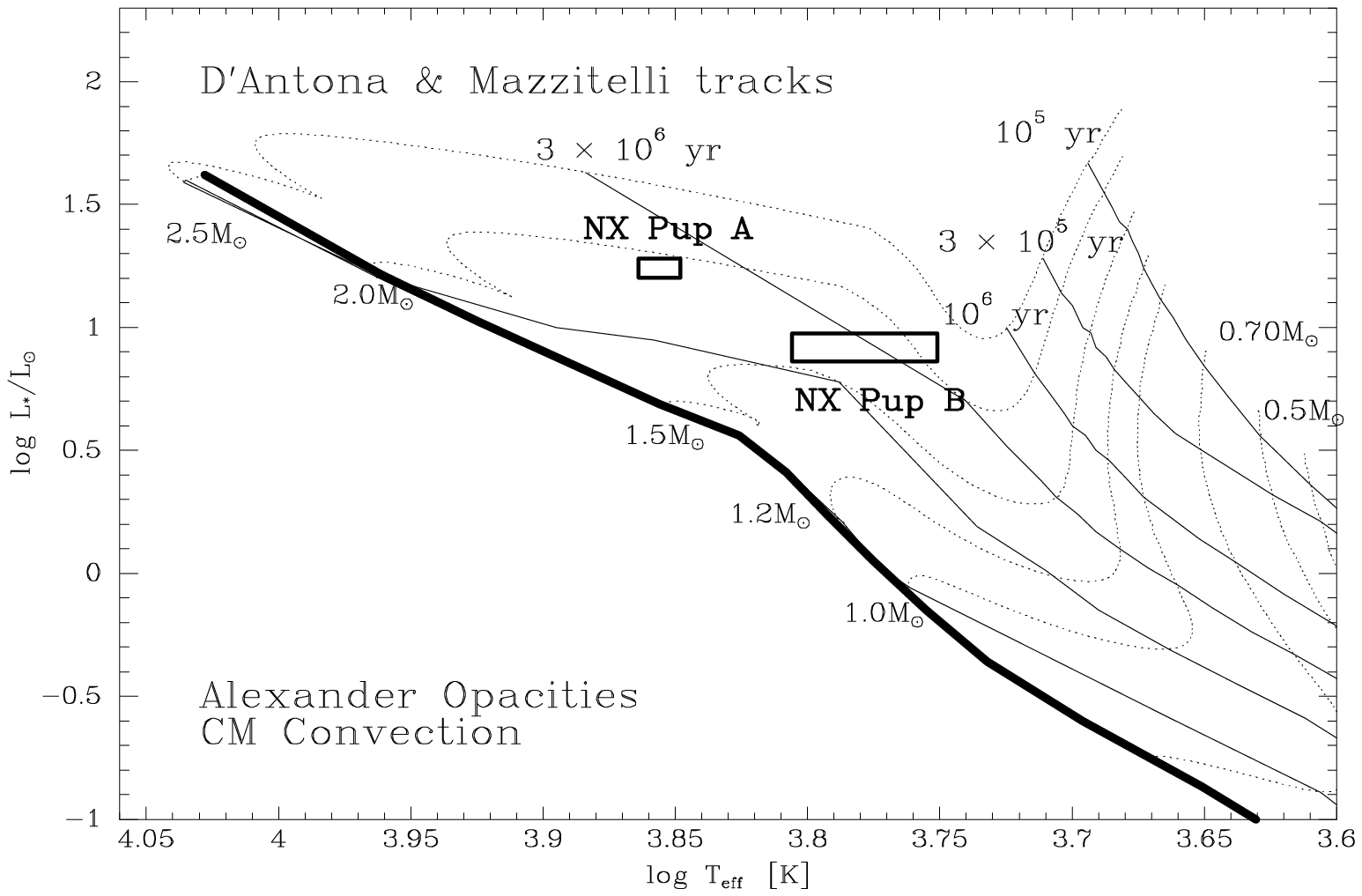,width=8.5cm,height=5.5cm}
                  \psfig{figure=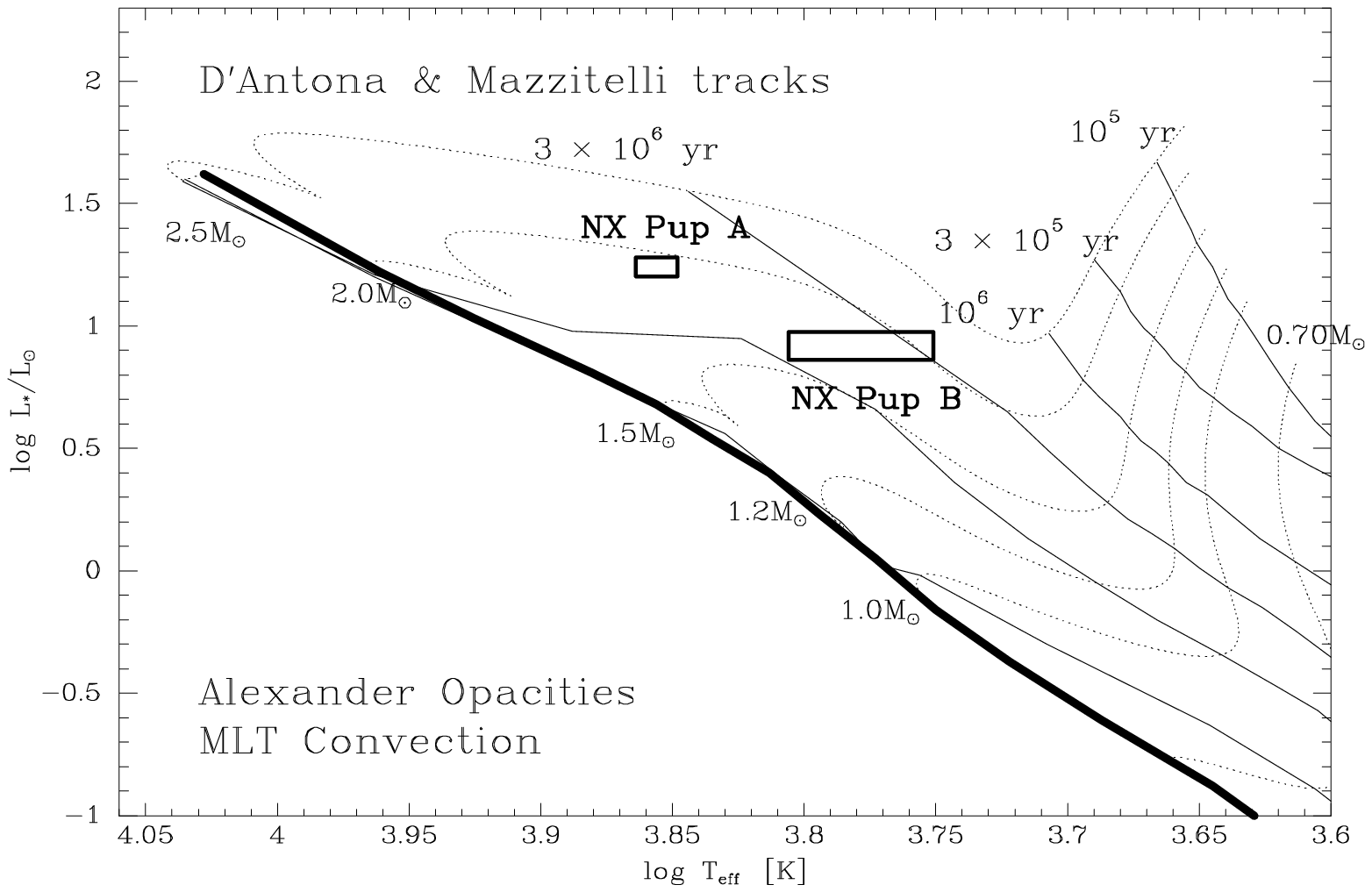,width=8.5cm,height=5.5cm}
                  \psfig{figure=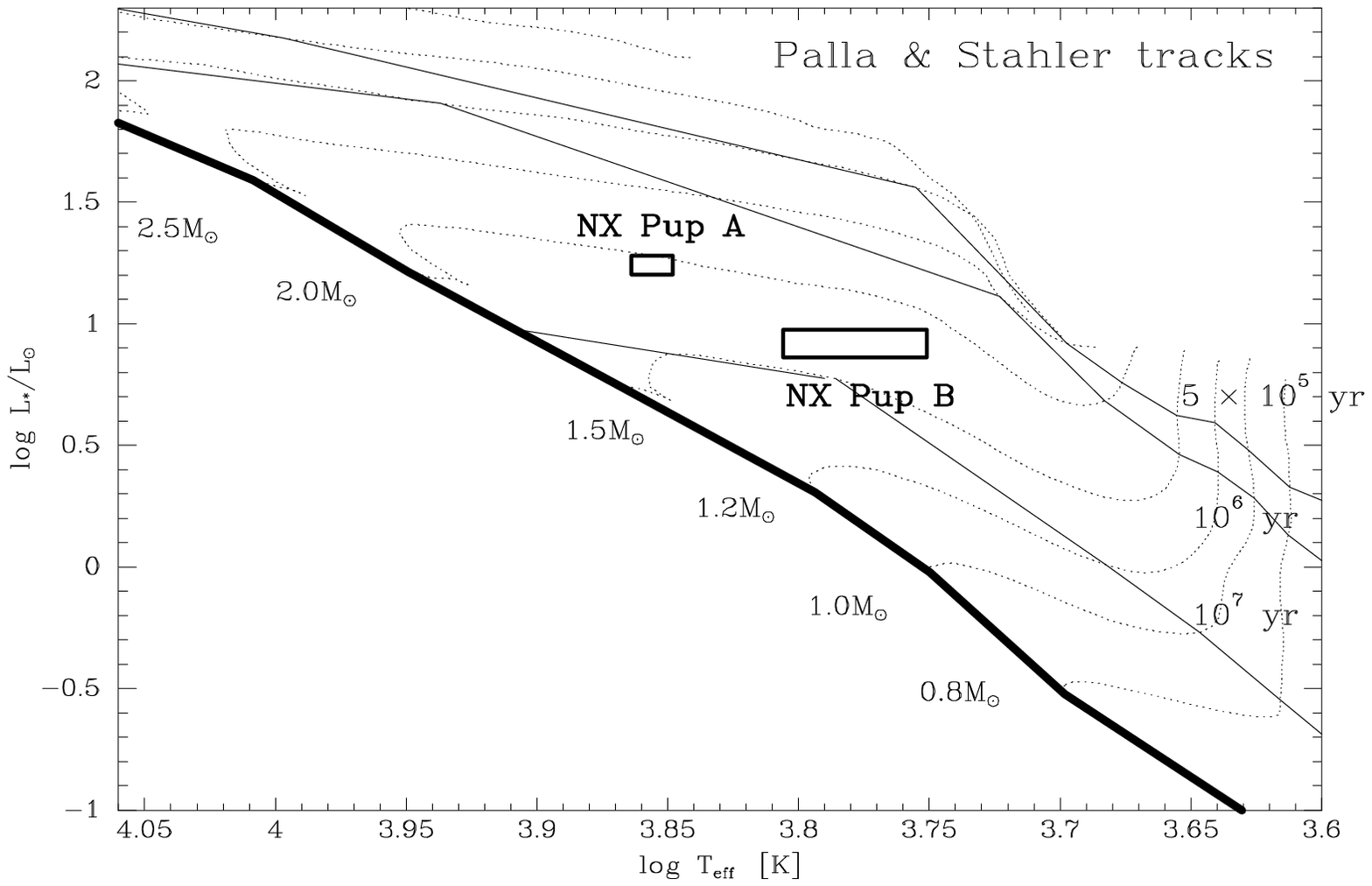,width=8.5cm,height=5.5cm}}}
\caption{\label{hrd}NX Pup A \& B placed on theoretical PMS evolutionary
tracks (dotted lines) based on different input physics. The solid lines
indicate isochrones and the Zero Age Main Sequence (ZAMS) is marked by
a bold solid line.
The various tracks are in good
agreement for intermediate mass stars (i.e.\ 1.5 M$_\odot$ to 2.5 M$_\odot$)
older than 10$^6$ yr.}
\end{figure}

\begin{table}
\caption{\label{evol}Evolutionary status of NX\,Pup\,A \& B}
\begin{center}
\begin{minipage}{70mm}
\begin{tabular}{ccc}
NX\,Pup    & A    & B \\
\hline
separation&\multicolumn{2}{c}{$0\farcs 126\pm 0\farcs 003$\footnote{11.3.1995}}\\
PA        &\multicolumn{2}{c}{$62\fdg 8\pm 1\fdg 7^a$}\\
SpT       &F0--F2 \footnote{Brand et al. 1983, Reipurth 1983, Blondel \& 
Tjin A Djie 1994} & F7--G4\\
L/L$_\odot$ &16--19 & 7--9 \\
mass &$\approx$\,2 M$_\odot$ &1.6--1.9 M$_\odot$  \\
age       &$3-5 \times 10^6$ yr & $2-6 \times 10^6$ yr \\ 
\end{tabular}
\end{minipage}
\end{center}
\end{table}

By assuming the same extinction for NX Pup B, its V--R colour
yields a spectral type F7V. However, the fact that NX Pup B's IR
excess is significantly smaller than that of NX Pup A might indicate
that while it suffers the same foreground extinction as NX Pup A,
its circumstellar extinction might be considerably less.	Studies by
Krautter (1980) and Brandt et al.\ (1971) indicate that the amount of 
foreground extinction in the direction of the Gum nebula might be as small
as A$_v$ $\approx$\,0\fm15 out till 500 pc from the Sun. If NX Pup B suffered
% Hans meinte, als Eigenname wuerde man "Sun" gross schreiben.
no additional extinction, its V--R colour would yield a spectral type of G4V. 
Accordingly, we compute
L$_{bol} \approx$\,7.3 -- 9.4 L$_\odot$. 

Mass and age determinations based on different sets of PMS evolutionary tracks
yield similar results. We used
theoretical tracks computed by D'Antona \& Mazzitelli (1994) and
Palla \& Stahler (1993), based on different input physics (i.e., opacities
and convection models), which
are in good agreement for intermediate mass
stars older than 10$^6$ yr (cf.\ Fig.\ \ref{hrd}). The results are summarized 
in Table \ref{evol}.

What is the relation of NX Pup A \& B to CG1? 
Based on molecular line observations with the SEST
Harju et al.\ (1990) show that the kinetic energy within
CG1 is too large for the globule to be gravitationally bound. They 
estimate a dynamical age of the order of 10$^6$ yr for CG1. 
Furthermore, they argue that
the enhanced [HCN]/[HNC] ratio in the head of the globule indicates
that a shock front reached CG1 and changed the chemistry within
the head. In total 75\,\% of all cometary globules in the Gum nebula
show indications of a shock induced change in their chemistry (presence
of ammonia, cf.\ Bourke et al.\ 1995) compared to only 40\,\% of all Bok globules
surveyed outside the Gum nebula. Bourke et al.\ further suggest that
the large percentage of ammonia detections and the fact that 50\,\% of
the CGs have an associated IRAS source is a strong indication that star 
formation is enhanced in CGs compared to Bok globules.

Harju et al.\ (1990) propose that CG1 and the nearby CG2 might just represent
sparse left overs of a single more massive molecular cloud which got slowly
dispersed due to photoevaporation. The evaporation was very likely
triggered by UV radiation from hot massive stars
located near the centre of the Gum nebula.
NX Pup A \& B may have formed some 3 -- 5 Myr ago within this
larger single molecular cloud, and hence considerably before CG1 reached its
current morphology.

The IR excess of NX Pup A can be approximated by the viscous accretion disk
spectrum, first modeled by Lynden--Bell \& Pringle (1974),
in which F$_\lambda$ falls off $\propto \lambda^{-4/3}$
towards longer wavelength. This rather simple model consists of a
thin viscous accretion disk made up of
concentric isothermal rings radiating like a black body.
We note that an accurate modeling of a physical accretion disk, especially
one that is also reprocessing stellar radiation, however, is much more complex 
and therefore it is not possible to derive parameters of a disk from
the spectrum alone (see e.g.\
Sonnhalter et al.\ 1995). The non-detection of 1.3 mm dust continuum emission
(Henning et al.\ 1994) suggests that the disk around NX Pup A is cut off
at about 20 AU possibly due to the presence of component B
(cf.\ Jensen et al.\ 1996 for a submm study of PMS binaries and
Artymowicz \& Lubow 1994 for model calculations).
For component B a single black body with a temperature of $\approx$\,1600 K
gives only a crude fit to the observed SED in the NIR. 
This could mean that there
is much less circumstellar matter left around NX Pup B compared to NX Pup
A, and that the distribution of matter
around NX Pup B is rather inhomogeneous because of the tidal torque of NX Pup A.

Clearly, for both NX Pup A and B more detailed model calculations and
spatially resolved imaging towards longer wavelengths are necessary to
put better constraints on the structure and geometry of the circumstellar
material around each star. Furthermore, high spatial resolution observations
of NX Pup near its minimum brightness would be interesting. One would
expect that NX Pup A then is completely obscured and would only be visible
in the optical through light scattered by circumstellar material,
similar to the active Herbig Ae/Be star Z CMa, where the IR companion
was detected in the visual by Barth et al.\ (1994) and Thi\'ebaut et al.\
(1995). Measurements of
the polarization of NX Pup A \& B near minimum brightness would be a test
for the scattered light hypothesis.

\section{Summary}

We have analysed quasi-simultaneous high angular resolution optical speckle 
and NIR adaptive optics data of the close Herbig Ae/Be binary star NX Pup. 
Within the observational errors both components appear to be coeval (3--5 Myr)
and have masses around 2 M$_\odot$ and 1.6--1.9 M$_\odot$, respectively. 
NX Pup appears to be older than the current morphological configuration
of CG1, which has a dynamical age of only 1 Myr.
The circumstellar
matter around NX Pup A can be described by a viscous accretion disk out
to a distance of $\approx$\,20 AU from the star. The outer part of the disk,
however, seems to be cut--off by the gravitational influence of the secondary.
The secondary itself has only a small amount of circumstellar matter left.

\acknowledgements
We are grateful to J.F.\ Claeskens for providing part of his
telescope time at the D1.54m. We thank the LTPV project for making their
data available via CDS. We thank P.\ Bouchet for communicating the list of
photometric ISO standards to us.
WB \& TL were supported by student fellowships of the European Southern
Observatory. WB acknowledges support from the Deutsche Forschungsgemeinschaft 
(DFG) under grant Yo 5/16-1. HZ acknowledges support from the DARA under grant 
05 OR 9103 0.
This research has made use of the Simbad database,
operated at CDS, Strasbourg, France, NASA's Astrophysics Data System (ADS),
and the IRAF PACKAGE C128, developed by E.\ Tessier at the
Observatoire de Grenoble.

\end{document}